\documentclass{article}

\usepackage{microtype}
\usepackage{graphicx}
\usepackage{booktabs} % for professional tables
\usepackage{enumitem}
\usepackage{hyperref}

\usepackage[accepted]{icml2023}

\usepackage{amsmath}
\usepackage{amssymb}
\usepackage{mathtools}
\usepackage{amsthm}

\usepackage[capitalize,noabbrev]{cleveref}

\theoremstyle{plain}

\theoremstyle{definition}

\theoremstyle{remark}

\usepackage[textsize=tiny]{todonotes}

\icmltitlerunning{}

\usepackage{caption}
\usepackage{subcaption}
\usepackage{graphicx}
\usepackage{wrapfig}
\usepackage[export]{adjustbox}
\usepackage{color}

\begin{document}
\twocolumn[
\icmltitle{Trust and ethical considerations in a multi-modal, explainable AI-driven chatbot tutoring system: The case of collaboratively solving Rubik’s Cube}




\begin{icmlauthorlist}
\icmlauthor{Kausik Lakkaraju}{sch}
\icmlauthor{Vedant Khandelwal}{sch}
\icmlauthor{Biplav Srivastava}{sch}
\icmlauthor{Forest Agostinelli}{sch}
\icmlauthor{Hengtao Tang}{sch}
\icmlauthor{Prathamjeet Singh}{sch}
\icmlauthor{Dezhi Wu}{sch}
\icmlauthor{Matt Irvin}{sch}
\icmlauthor{Ashish Kundu}{yyy}

\end{icmlauthorlist}

\icmlaffiliation{sch}{University of South Carolina, Columbia, South Carolina, USA}
\icmlaffiliation{yyy}{Cisco Research, San Jose, California, USA}

\icmlcorrespondingauthor{Kausik Lakkaraju}{kausik@email.sc.edu}

\icmlkeywords{Machine Learning, ICML}

\vskip 0.3in
]



\printAffiliationsAndNotice{} % otherwise use the standard text.

\begin{abstract}
Artificial intelligence (AI) has the potential to transform education with its power of uncovering insights from massive data about student learning patterns. However, ethical and trustworthy concerns of AI have been raised but are unsolved. Prominent ethical issues in high school AI education include data privacy, information leakage, abusive language, and fairness. This paper describes technological components that were built to address ethical and trustworthy concerns in a multi-modal collaborative platform (called ALLURE chatbot) for high school students to collaborate with AI to solve the Rubik’s cube. In data privacy, we want to ensure that the informed consent of children, parents, and teachers, is at the center of any data that is managed. Since children are involved, language, whether textual, audio, or visual, is acceptable both from users and AI and the system can steer interaction away from dangerous situations. In information management, we also want to ensure that the system, while learning to improve over time, does not leak information about users from one group to another.
\end{abstract}

% \textbf{Practitioner Notes}

% \textbf{What is already known about this topic}

% \begin{itemize}
    
% \item Artificial intelligence (AI) has transformative potential in reshaping education.
% \item AI has afforded a variety of innovations in education such as personalized learning systems, automated assessment systems, and intelligent systems with capacity of facial/speech/emotion recognition.
% \item In each of the AI techniques used, there are reported trust and trustworthiness issues.

% \end{itemize}

% \textbf{What this paper adds} 
% \begin{itemize}
% \item This paper introduces an AI system that can collaborate with students to solve the Rubik's cube. 
% \item This paper reports how a combination of innovative methods in design, implementation, and evaluation are being used to make the AI system trustworthy.
% \item This paper specifies measures on protecting user information and avoiding ethical issues about AI for K-12 education.
% \end{itemize}

% \textbf{Implications for practice and/or policy}
% \begin{itemize}
% \item The paper presents  a general design and implementation approach for handling trust issues with an educational AI system for STEM learning that could be broadly applicable to areas in which humans must collaborate with AI for education or work.
% \end{itemize}

% -------------------------
\section{Introduction}
Artificial intelligence (AI) has been used to automate time-consuming tasks, solve problems that many humans struggle to solve, as well as achieve superhuman performance on problems that humans have been studying for centuries. However, for the vast majority of modern successes, the strategies that the AI agent learns is a ``black-box'' that is not transparent to humans. Furthermore, the black-box problem is also present when we want to express our own strategies to an AI agent as most human knowledge cannot easily be communicated to modern AI agents. As a result, an explicit instructive and productive form of \textit{collaboration}, the positive feedback loop of humans learning from AI and AI learning from humans, is blocked by the problem of the black-box.

Explainable AI (XAI) \citep{adadi2018peeking} methods seek to allow information to flow from AI to humans. XAI is an open field where even the definition of explainability is debated \citep{lipton2018mythos}. The particular methods used to make an AI system explainable depend on the AI algorithms used, the application, and the audience. In this particular case, we seek to use deep reinforcement learning \citep{sutton2018reinforcement} and inductive logic programming \citep{muggleton1991inductive}, to find solutions to the Rubik's cube, that a high-school student can easily understand. Furthermore, we seek to make these explanations \textit{personalized} in terms of both the explanations themselves as well as the manner in which they are communicated.

\begin{wrapfigure}{r}{0.26\textwidth}
  \begin{center}
    \includegraphics[width=0.28\textwidth]{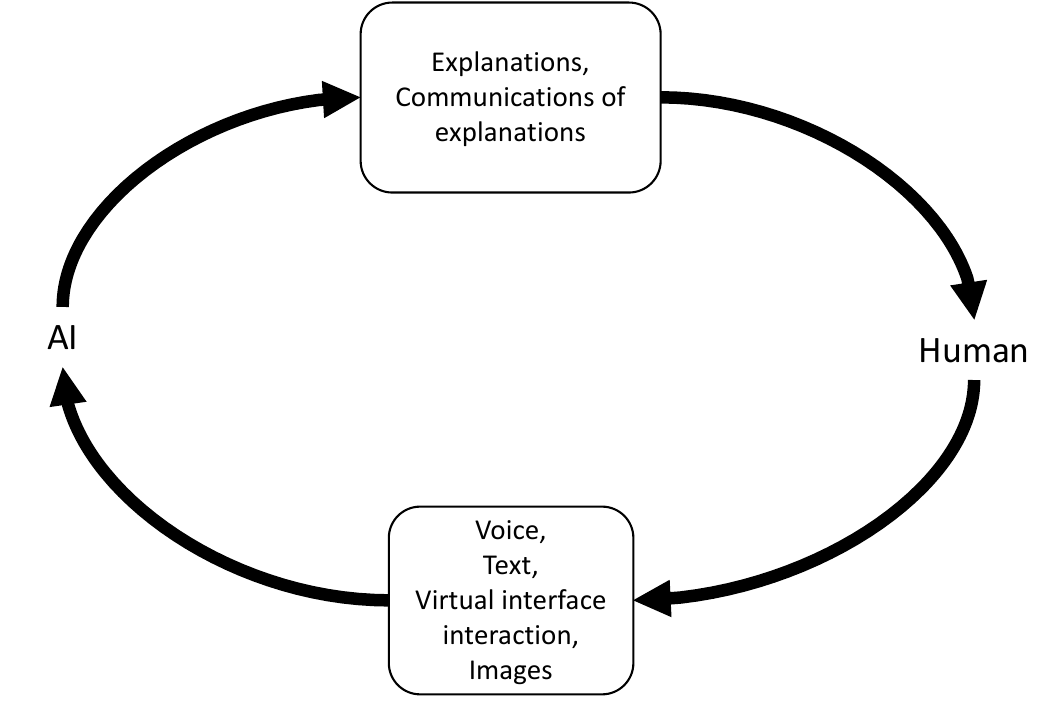}
  \end{center}
  \caption{The information that facilitates collaboration between human and AI that needs to be protected from leakage.}
\label{fig:inforleak}
\end{wrapfigure}

% -----
Human computer interaction methods can be used to allow information to flow from humans to AI. These methods include speech recognition, natural language processing, visual interfaces, and image recognition. Using speech recognition and natural language processing, a user can communicate with AI both through their voice and through text. In our setting, these communications can express their own ideas about how to solve the Rubik's cube or can ask questions about a particular explanation. Using a visual interface that  employs image recognition, the user can also express their own ideas about how to solve the Rubik's cube or ask the AI system about a particular configuration that they would like to solve.

Both the AI-to-human and human-to-AI communication directions have personal information that must be protected against information leakage. This includes: (1) explanations; (2) communications of explanations; (3) user voice data; (4) user text data; (5) visual interface usage data; and (6) user image data. An overview of this is shown in Figure \ref{fig:inforleak}. While the need to protect against information leakage is ubiquitous in interaction with digital devices, the type of data produced when interacting with AI presents unique questions for understanding what this personal information contains and how to protect it. For example, seemingly innocuous information, such as two users finding different explanations or two users finding the same explanation communicated in different manners, could reveal information related to learning styles that a user may want to keep private.

Multiple major human stakeholders in the environment interact with the AI system. Foremost is the {\em student learner}. Next are the {\em parents} who are tracking the student's progress and the {\em teacher} who is responsible for the student's learning progress. Finally are the {\em student peer} who studies along with a student and can help improve learning and {\em school administrator} who sets policies for students learning. So, although student learner is the primary human user group, there are at least 4 other stakeholders that may look at the AI's output, seek details, including explanations of the system's behavior and learning impact. 

We focus on two aspects on ethics in the AI-driven chatbot design for solving Rubik's Cube problems: 
(1) acceptable conversations and (2) preventing information leakage.  For acceptable conversations, we consider preventing abusive language and maintaining suitable conversation complexity (style). In information leakage, we consider not leaking history of a single student's learning as well as comparative learning of pairs or larger groups of students without suitable reason or permission.

% -------------------------
\subsection{Case Study}

% The \textit{Rubik's cube} is a three-dimensional puzzle invented in the year 1974. 
We chose the Rubik's Cube puzzle as our problem-solving tasks in that it is inherited with complexity, difficulty and experiential learning value for children to solve hard problems mathematically and spatially that requires curiosity and patience. We begin by formalizing the terminology used. A cube consists of 54 stickers, and each sticker is a colored square with any of the six colors: red, green, blue, orange, white, and yellow. The Rubik's cube consists of 6 faces, where each face has 9 stickers. Each sticker is on a ``cubelet'', where cubelets are smaller cubes within the Rubik's Cube. Based on the sticker count, these cubelets are classified as corner, edge and center cubelets. Corner cubelets have three stickers, edge cubelets have two stickers, and center cubelets have one sticker. There are six center cubelets, 12 edge cubelets, and eight corner cubelets. Each of the cube faces can be rotated 90 degrees clockwise or counterclockwise. The cube can be moved in 12 different ways where each of the six faces can be rotated either clockwise or counterclockwise. 
%These moves are represented with letters that denote the face being moved where F, B, L, R, U, and D represent the front, back, left, right, up, and down faces, respectively. These letters represent a 90 degrees clockwise rotation of the face while these letters followed by an apostrophe represent move in the counterclockwise direction (i.e., F'). We call these moves ``atomic actions.''

% \begin{wrapfigure}{r}{0.55\textwidth}
\begin{figure}
  \begin{center}
    \includegraphics[width=0.25\textwidth]{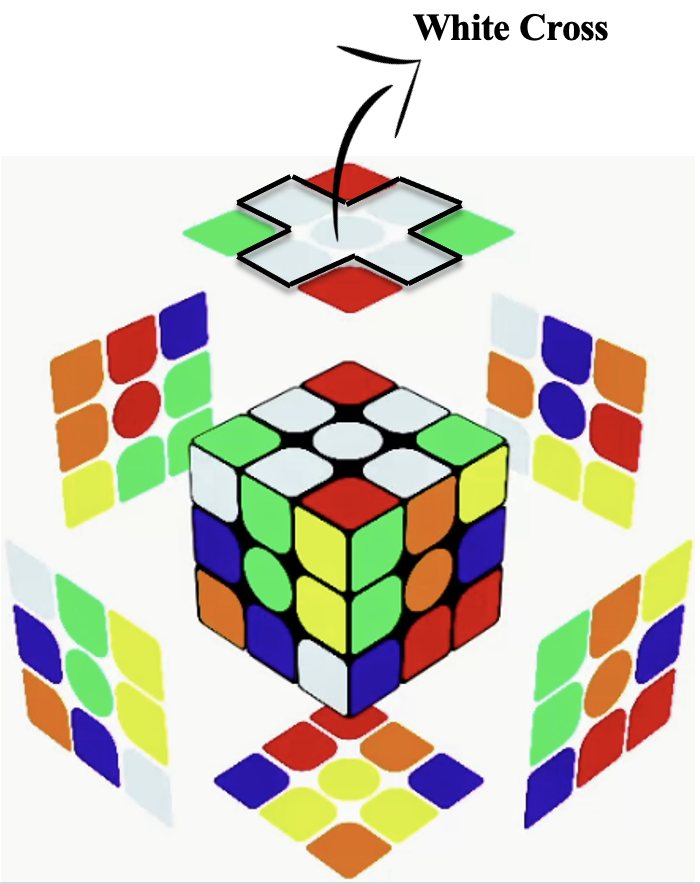}
  \end{center}
  \caption{The Solved white cross}
  \label{fig:whitecross}
\end{figure}
% \end{wrapfigure}

In this case study, we focus on one specific configuration that called the {\em white cross} as shown in Figure~\ref{fig:whitecross}. This was the same use-case that was used in \cite{allure}. We focus on this configuration because, in many tutorials for solving the Rubik's cube, the white cross is one of the initial sub-goals \citep{ruwix}. An example of part of an explanation for the solving of white cross is shown in Figure \ref{fig:flowsteps}. This figure shows the scrambled state of the Rubik's cube where the white-orange edge cubelet is not in place, where the white side is aligned to the orange center cubelet, and the orange side is aligned to the yellow center cubelet. %First, to align the orange side of the edge cubelet to the orange center cubelet, we move the edge cubelet away by performing move D', then align the orange side of the edge cubelet by performing F', which moves white-green edge cubelet out of place. Then, to move the white-orange edge cubelet in place, we perform move R, and then perform move F to put the white-green edge cubelet back in place.

% -------------------------
\subsection{Problem Scope}

In this research study, our major goal is to improve the AI-driven interactive multi-modal chatbot (or bot for short) infrastructure called ALLURE which was built to engage high school students to learn how to partner with the AI for collaboratively solving Rubik's Cube problems. The student uses the ALLURE interactive environment \cite{allure} to learn to play Rubik's Cube (RC) games and improve their mathematical thinking and spatial skills over time. It is a collaborative virtual assistant (chatbot) built using natural language processing (NLP) based on RASA framework \cite{bocklisch2017rasa}. 
% and image recognition techniques that can interact with the user to understand their gaming and learning intents, track progress and manage interactions. The users can refer to previous conversations as part of their context to initiate a game and the system can  track user's conversation and assess emotions like sentiments to detect frustration, user engagedness and learning progress. ALLURE is embedded with an explainable RC solver, DeepXube, that proposes a solution to the goal configuration and an explanation. The system translates explanations from first order logic to natural text using templates that can be customized to different emotional states of the users. 
Furthermore, the chatbot allows users to communicate their own ideas for subgoals using a virtual cube that they can edit. Users can communicate their ideas for algorithms by providing the system with examples of the algorithm and having the ILP system induce a logic program, which will then be translated to natural language for verification. Alternatively, the user can directly communicate their algorithm using natural language, with the system asking for clarification when encountering any ambiguities.

In ALLURE, we aim to address two pressing and significant research challenges to effectively connect education with transformative AI technologies. (1) The trust problem in ALLURE is to maintain a safe environment for children and to transparentize the black-box communication processes inherited from the collaborative AI algorithms in the Rubik's Cube learning environment. 
% Specifically, we want to ensure that informed consent of children or parents, and teachers, are at the center of any data that is managed. 
Since children are involved, it is critical to ensure that language, whether textual, audio or visual,
is acceptable both from the users and the AI, and the system is able to steer interaction away from dangerous
situations. (2) In terms of information management, we also ensure that the system,
while learning to improve over time, does not leak information about users from one group to another for the sake of privacy and security purposes.

% However, we recognize the need for careful user evaluation to determine the effectiveness of these methods for different user groups given their susceptibility to trust issues like bias \cite{chatbot-trust,sentiment-bias}. 
% -------------------------
% -------------------------

% -------------------------
% -------------------------
\section{Literature Review}
\subsection{Collaborative Virtual Assistants, i.e., Chatbots}

% \begin{figure}
%     \centering
%     \includegraphics[width=0.8\textwidth]{figs/generic-chatbot.pdf}
%     \caption{Architecture of a data-driven collaborative assistant.}
%     \label{fig:generic-chatbot}
% \end{figure}

The core technical problem in building chatbot systems is that of dialog/dialogue management (DM), i.e., creating dialog responses to user's utterances. Given the user's utterance, the system analyzes it to detect user's intent and employs a policy, which is a mapping between state observed and action to perform and produces automatic responses through computational selection and generation from a library. 
The common approach to create dialog responses is to maintain a list of supported user's intents and the corresponding pre-canned responses (policy library). This is often the first and fastest approach to introduce a chatbot in a new application domain. 

% However, sophisticated task-oriented chatbots need to apply advanced natural language processing (NLP) methods, reason with uncertainty and optimize for dialog goals to produce a policy with possibilities to integrate available data sources.
The system architecture of a typical data-consuming DM is shown in Figure~\ref
%{fig:generic-chatbot} 
{fig:controlflow}, in which the language understanding module (LU) processes the user utterance and the state of dialog is monitored (by ST module). The strategy to respond to user's utterances is created with reasoning and learning methods (PG).
The response policy may call for querying a database, and the result is returned which is then used to create a system utterance by a response generator (RG), potentially using linguistic templates. The response policy may alternatively invoke a specialized engine like a game solver to create solution content that is returned as part of the response.
The policy may also include the decision not to answer a request if it is unsure of a query's result correctness.

Note that the DM may use one or more domain-independent resources like user models, word embeddings (representations)/ language models, as well as language resources like lexicons from third party services. Furthermore, they may use one
or more domain-specific data bases (sources). These third-party resources can be the first source of bias. Another source of bias can be the user's own input that the force a system blindly processing it to generate a biased response. The third source of bias can be the specialized solvers and information sources that the chatbot relies on for its response.  

% Common chatbots use static domain-dependent databases 
% like product catalogs or user manuals. The application scenarios become more  compelling when the chatbot works in a dynamic environment, e.g., with sensor data, and interacts with groups of people, who come and go, rather than only an individual at a time. 
When the domain is dynamic, 
%In such situations, 
the agent has to execute actions to monitor the environment, model different users engaged in conversation over time and track their intents, learn patterns and represent them, reason about best course of actions given goals and system state, and execute conversation or other multi-modal actions. As the complexity of the DM increases along with its dependency on domain dependent and independent data sources, the challenge of testing it increases as well. Testing for trustworthy behavior is adds another dimension to basic functional testing. 

% Given a database, it is imperative to understand a user’s intent by parsing her query and provide accurate information that satisfy her goal in a timely manner.
% To achieve this, we need a dialog manager (DM) that can manage its conversation with the user and can search the database efficiently.

% -------------------------
\subsection{Ethical and Trustworthy Concerns about AI in Education}

\cite{ai-k12} summarized four ethical concerns of AI being used in education, including privacy, surveillance, autonomy, and bias and discrimination. Specifically, privacy concerns about AI mainly address the situation that students’ information, including their personal demographics, preferences, and performance metrics about their learning progress, are accessed and retained by others (including schools, third-party service providers, and even the public) (\cite{Regan_Steeves_2019}, \cite{blair2020effect}).
Surveillance concerns mainly describe the monitoring and tracking system of student learning activities embedded in the AI system \cite{1886}. Agency concerns denotes the issue that recommendations or even selections made by the AI system may transcend learner agency to direct their own learning progress \cite{Piano2020EthicalPI}. Then the bias and discrimination concerns mainly result from the existing bias or unequal power structure such as gender bias which has been ingrained in the AI algorithms \cite{ai-k12}.
We focus on privacy and language (fairness) issues that we discuss next.

% -------------------------
\subsection{Privacy Concerns with AI Services}
\label{sec:priv-rw}

Protecting students’ privacy is a critical task because of the increase in the offering of online courses on various platforms. 
% Privacy concerns mainly address the situation where students’ information, including their personal demographics, preferences, and performance metrics about their learning progress, are accessed and retained by others (including schools, third-party service providers, and even the public) (\cite{ai-k12}). 
As our education field increasingly realizes the detrimental effect of privacy violations, students are usually asked to complete consent forms which may partly mitigate this issue. However, the authors of \cite{ai-k12} pointed, students are usually not left with many choices when the school or the learning platform that the school adopts make sharing data with the platform mandatory. In addition, we cannot assume that each student is clearly aware of what sharing personal data means and the potential consequence resulting from sharing their personal data at the point of giving the consent (\cite{ai-k12}).

With chatbots, one specific privacy issue of concern is information leakage. \cite{ethical-dialog} describes an information leakage framework that can recover user's sensitive information from conversation. This issue involves ensuring that information given by users to a chatbot is not released, even inadvertently, to other users of the same chatbot or the same platform. 
% This is further complicated by the fact that over time,
% a chatbot may get personalized to a user’s needs but the person
% may not want to share their personalized information with the
% developers. Moreover, shared information may spread when
% other users interact with a person over time.  
In our setting, we focus on game performance that one student (player) may not want to share with another.

% -------------------------
\subsection{Chatbot Conversation Style Issues and Dialog Generations with AI Services}

%There is growing awareness in the research community about fairness issues with AI services. Here, we focus on concerns related to the linguistic style of conversation and the dialog setting.

% -------------------------
\subsubsection{The Language of Conversation}

The concern with language of a chatbot is that it should not respond with hateful or abusive
language, and converse in a style that is appropriate to the user.

There is a growing body of work to detect hate 
speech (\cite{hateoffensive}) and abusive language (\cite{twitter-curse}) online using words and phrases which people have annotated. The authors in the former paper define hate speech  as {\em language that is used to expresses hatred towards a targeted group or is intended to be derogatory, to humiliate, or to insult the members of the group}. Their checker, which is publicly available, has a logistic regression with L2 regularization to achieve automatic detection of hate speech and offensive language. 
% In studying abusive language online, the authors in \cite{twitter-curse} explore the prevalence of cursing on Twitter which serves as a platform for utterance and conversation. They found that people curse more online than in physical environment, among same gender, when they are angry or sad, as their activities increase during the day, and when in relaxed or formal environments. 

The concern over conversation style and complexity has to do with making sure that AI services interact with users in the most useful and seamless way. If a chatbot responds to user's questions with a terminology that the user is not familiar with, the user will not get the required information and will not be able to solve the problem at hand. 
%So it is important that a dialog system uses a language that the user can understand and find useful for the purposes of the interaction. 
\cite{dialog-complexity} propose a measure of \textit{dialog complexity} to characterize  how participants in a conversation use words to express themselves (utterances), switch roles and talk iteratively to create turns, and  span the dialog. 
% They measure the complexity of service dialogs at the levels of utterances, turns and overall dialogs. 
The method takes into consideration the concentration of domain-specific terms as a reflection of user request specificity, as well as the structure of the dialogs as a reflection of user's demand for (service) actions. Their checker can be used as an additional component to improve ALLURE.

% -------------------------
\subsubsection{Dialog Generation}

Chatbots can be fraught with ethical issues. An extreme %and anecdotal 
example is the Tay Twitter chatbot (\cite{tay,tay-comments}), released by Microsoft in 2016, that was designed to engage with people on open topics and learn from feedback, but ended up getting manipulated by users to exhibit unacceptable behavior via its extreme responses. But users may not want the chatbots they are interacting with to exhibit the same behavior, especially when the users are children. 

One of the first papers to look at ethical issues in dialog systems was  \cite{ethical-dialog} which the authors described implicit biases, adversarial examples, potential sources of privacy violations, safety concerns, and results reproducibility.
More recently, \cite{dialog-ethics-2020-queens} looked at gender related fairness issues in dialogs while \cite{dialog-ethical-recipes} proposed a set of techniques to control problematic behavior of dialog systems including sometimes not answering or changing the conversation's subject when the chatbot is unsure about the ethical ramification of its response.

% -------------------------
% -------------------------
\section{Methodology}

\subsection{Finding Explainable Solutions to the Rubik's Cube Game}

We formally define an explanation as a set of macro actions, also referred to as ``algorithms'' by the Rubik's cube community, that can be composed by a human to solve any instance of the Rubik's cube. Each macro action is a sequence of atomic actions that is associated with preconditions and effects. To ensure that these macro actions are suitable to humans, the preconditions and effects should be simple and concise and the sequence of atomic actions should not be too long. Furthermore, the number of macro actions should be small enough so that someone can easily memorize them. To discover these macro actions,  deep reinforcement learning and search algorithm capable of solving the Rubik's cube, DeepCubeA \citep{agostinelli2019solving}, with inductive logic programming (ILP) \citep{muggleton1991inductive} were combined together. We first give a brief overview of DeepCubeA. 
% and the ILP method we use. 
Then, we describe our algorithm for finding explanations in the form of logic programs for solving the Rubik's cube, which we call DeepXube. Finally, we describe how we translate these logic programs to English using natural language generation techniques \citep{mcdonald2010natural}.

\subsubsection{DeepCubeA}
DeepCubeA \citep{mcaleer2018solving, agostinelli2019solving}, is an artificial intelligence algorithm that uses deep reinforcement learning \citep{sutton2018reinforcement} to train a deep neural network (DNN) \citep{schmidhuber2015deep} to map a Rubik's cube configuration to the estimated number of steps it will take to solve the cube. This DNN is then used as a heuristic function for A* search \citep{hart_1968}, which finds the sequence of steps needed to solve a given configuration of the Rubik's cube. %Through manual inspection of the DeepCubeA algorithm, one could obtain some vague information about how to solve the cube. For example, conjugate triplets (performing a move $a$, another move $b$, and then reversing the first move with $a'$) were frequently used by the algorithm. However, the knowledge of when to apply these conjugate triplets and what other types of strategies were needed to solve the cube was beyond the scope of the DeepCubeA algorithm.

\subsubsection{DeepXube}
We discover macro actions by first designating effects of a macro action that are deemed to be desirable by humans. These are also referred to as \textit{focused effects} \citep{allen2020efficient}. In the context of the Rubik's cube, we define a single focused effect: a cubelet is moved to its correct position without disturbing other cubelets that are already in their correct position. Given focused effects, we can then search for macro actions using the DeepCubeA algorithm. This is done by generating multiple configurations by randomly scrambling the Rubik's cube and then, for each configuration, using DeepCubeA to put a particular cubelet in the correct position without disturbing any other cubelets. To achieve this, we use hindsight experience replay \citep{andrychowicz2017hindsight} to train DeepCubeA to be able to achieve any configuration of the Rubik's cube, even partially specified configurations. Because we want macro actions that are not too complex, we assign a complexity score to each macro action based on the length of atomic actions it contains. We then prioritize less complex macro actions over more complex ones.

To learn preconditions, we select the macro action that has the lowest complexity score and use Popper \citep{cropper2021learning} to induce a logic program for the precondition of this macro action. If a precondition is successfully learned, the macro action and its precondition are added to a set of learned macro actions and preconditions. We then apply these learned macro actions to the randomly generated configurations, updating each configuration to which a macro action is applied to be equal to the configuration resulting from applying that macro action, until none of the configurations match the preconditions. This method ensures that we do not learn any additional macro actions unless we find that the current ones are insufficient. We then repeat this process until all the configurations are solved. %The entire DeepXube algorithm is shown in Algorithm \ref{alg:deepxube}.

% ---
%
%\begin{algorithm}[tb]
%	\caption{DeepXube}
%	\label{alg:deepxube}
%	\begin{algorithmic}[1]
%        \State {\bfseries Input: DeepCubeA solver, Focused effects $\mathcal{F}$, Background knowledge $BK$, ILP system}
%        \State Generate a set of problems $\mathcal{S}$
%        \State Initialize list of learned macro action and precondition tuples $\mathcal{MP} = []$
 %       \While {There is a problem $s$ in $\mathcal{S}$ that is not solved}
 %           \State Find macro actions $\mathcal{M}$, that result in a focused effect $f$ in $\mathcal{F}$, for states in $\mathcal{S}$ with DeepCubeA.
 %           \State Find macro action $m$ in $\mathcal{M}$ with the lowest complexity \Comment{i.e. by macro action length}
 %           
 %           \State Based on focused effects $F$, find positive examples $E^+$ and negative examples $E^-$ % $\mathcal{S}$ for $m$

%            \State Use ILP system and $BK$ to attempt to find precondition $p$ for $m$
 %           \If {$p$ learned successfully}
 %               \State append $(m, p)$ to $\mathcal{MP}$
 %               \While {There exists $(m,p)$ in $\mathcal{MP}$ where $p$ matches a subset of $\mathcal{S}$} %\Comment{Apply all macro actions to problems}
%                    \For {$s$ in matching subset}
%                        \State Apply $m$ to $s$ to produce $s'$
%                        \State Update $s$ in $\mathcal{S}$ to $s'$
%                    \EndFor
 %               \EndWhile

 %           \EndIf
            
 %       \EndWhile
%	\end{algorithmic}
%\end{algorithm}

% ---

\subsubsection{Natural Language Generation}

After learning logic programs for the preconditions of the macro actions, 
we next generate their plain English descriptions  
%using Natural Language Generation techniques \citep{nlg-suvey}. 
% we now generate plain English descriptions of these logic programs using Natural Language Generation techniques \citep{nlg-suvey}. 
%Specifically, we 
following a template based method from Natural Language Generation  \citep{nlg-suvey} that maps each logical predicate to an English sentence  and then  arranges (e.g., sorts) them according to  readability heuristics. 
Since we are working in education domain where children may prefer different linguistic styles, one could dynamically rephrase descriptions of the preconditions based on user feedback and leveraging powerful language models \citep{vaswani2017attention, devlin2018bert, brown2020language}. Note that these language models, despite their generality, have often been reported as a source of bias. 

% -------------------------
\subsection{Key Specialized Components in ALLURE}
\label{sec:components}

\begin{figure*}
{\small
    \centering
    \includegraphics[width=\textwidth]{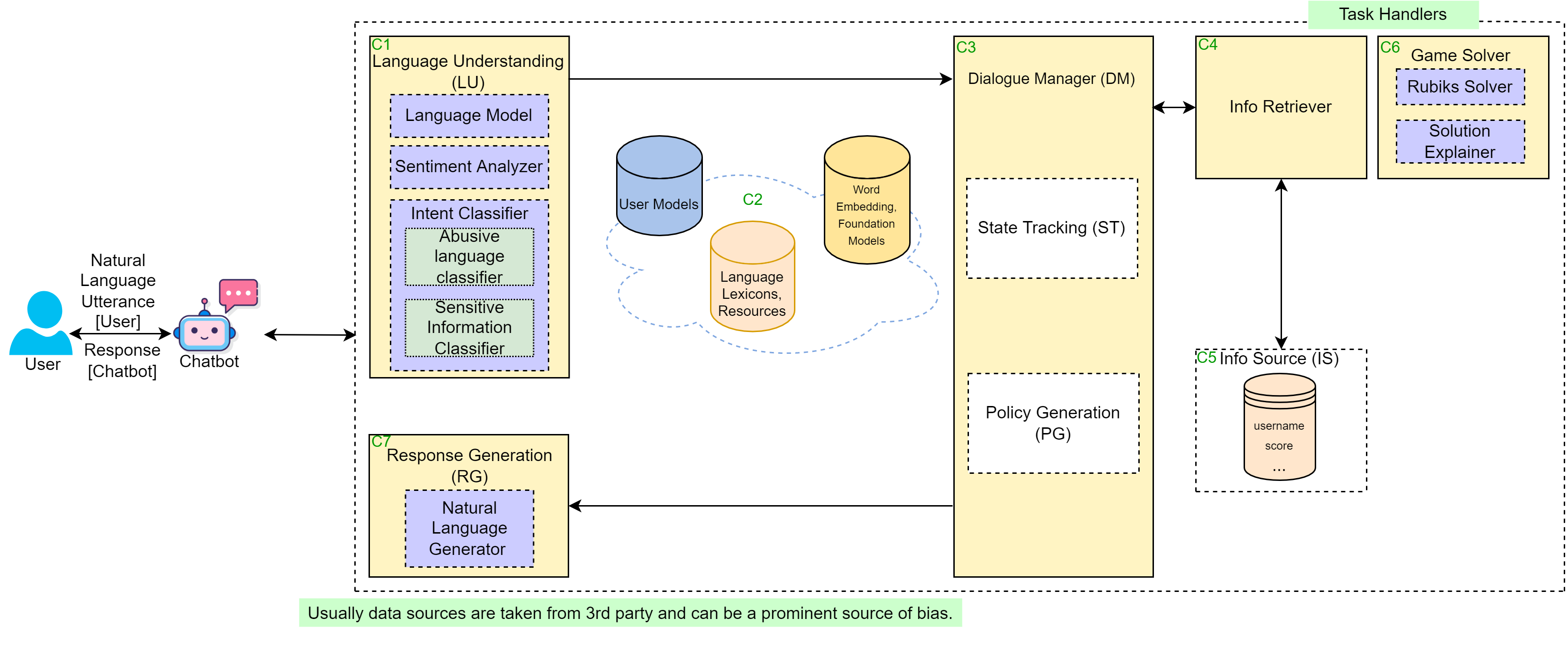}
    \caption{Architecture of a general data-driven collaborative assistant and our specialized  components.}
    \label{fig:controlflow}
}
\end{figure*}

RASA framework was used to build the ALLURE chatbot. Figure~\ref{fig:controlflow} shows the key components that the authors of \cite{allure} have used in ALLURE and the additional components we added (Sentiment Analyzer, Intent Classifier (modified), Natural Langauge Generator, Rubiks solver (using DeepXube) and Solution Explainer) to the system to prevent information leakage and detect sentiment and abusive language from user utterance.  
%(depicted in Figure~\ref{fig:controlflow})  
% below. 
\begin{enumerate}
   \item \textbf{Language Understanding (C1)}: This contains different sub-components which are responsible for Natural Language Understanding (NLU) in the system. 
   \begin{itemize}
       \item \textbf{Language Model (LM)}: This refers to the language model used in the conversational pipeline which is the Spacy Language Model provided by RASA.
       \item \textbf{Sentiment Analyzer}: This is a custom component we have built. It estimates the sentiment value of the user message. It uses the VADER \cite{hutto2014vader} sentiment analyzer to estimate the semantic orientation (positive, negative or neutral) and also estimates how positive, negative or neutral a sentiment is. 
       \item \textbf{Intent Classifier}: This component classifies the intent of the user message. We extended the capability of the intent classifier the authors of \cite{allure} have used to classify user utterances with abusive language and also to classify whether the user utterance has any sensitive information which has to be protected from being accessed by other users.
   \end{itemize}
   \item \textbf{User models, word embeddings and language lexicons (C2)}: User models, word embeddings (representations) and language lexicons will be used by the Dialogue Manager.
   \item \textbf{Dialogue Manager (C3)}: This monitors the state of the dialogue and also generates the policy for responding to the user utterances.
   \item \textbf{Info Retriever (C4)}: This component retrieves the relevant information from the database and passes it to the DM whenever the user asks for any information which is not sensitive. 
   \item \textbf{Info Source (C5)}: The database consists of all the user information that is required. This includes information like username, gender, score, games won and skill level of the user.  
   \item \textbf{Game Solver (C6)}: This contains 2 sub-components: Rubik's Solver and Solution Explainer.
   \begin{itemize}
       \item \textbf{Rubik's Solver}: It uses the user's current Rubik's Cube configuration to output the random states and solutions which will be passed to the Solution Explainer.
       \item \textbf{Solution Explainer}: The Solution Explainer will be trained on random states and solutions given by the Rubik's Solver to give out a set of predicates in the form of rules.
   \end{itemize}
   \item \textbf{Response Generation (C7)}: The response generated by the chatbot will depend on the user intent. For example, if the user uses abusive language, the chatbot will warn the user. When asked about the steps to solve for a particular pattern on the Rubik's cube, the Natural Language Generator (NLG) generates a response in human understandable language from a set of predicates produced by the Solution Explainer.
\end{enumerate}

% -------------------------
\subsection{Trustworthy Game Interaction}

In this section, we make an argument on why or how ALLURE chatbot system becomes trustworthy and reliable after adding our proposed components. 
The chatbot assesses the sentiment of user: negative, positive or neutral. Based on the estimated sentiment, the chatbot will respond accordingly. For example, the user might feel down or discouraged sometimes while solving the Rubik's cube. When that happens, the bot will try to encourage the user to keep solving or to take a short break. If the user utterance contains any abusive text/speech, the bot responds to the user with incremental warnings. This will discourage the user from using abusive language in the future.
If the user asks for sensitive information about other users, the chatbot will deny the user request by saying that the information that the user is seeking is confidential and hence cannot be shared. This makes the bot reliable in terms of keeping the user information safe and secure. It also prevents the system from leaking any sensitive information.

% These concepts are illustrated in the form of a conversation in Section 5.1.

% -------------------------
\section{Discussion}

% -------------------------
\subsection{A Complete Example}
\label{sec:example}
In this section, we will be discussing an entire conversation between the bot and the user to solve the white cross for the given configuration. We are taking the similar example of the white cross as shown in Figure \ref{fig:flowsteps}. Here, we want to highlight the major functionalities that would improve the ALLURE chatbot:

\noindent 1. Explainable solutions

\noindent 2. Handling use of improper language

\noindent 3. Handling information leakage

\begin{figure*}
    \centering
    \includegraphics[width=0.8\textwidth,left]{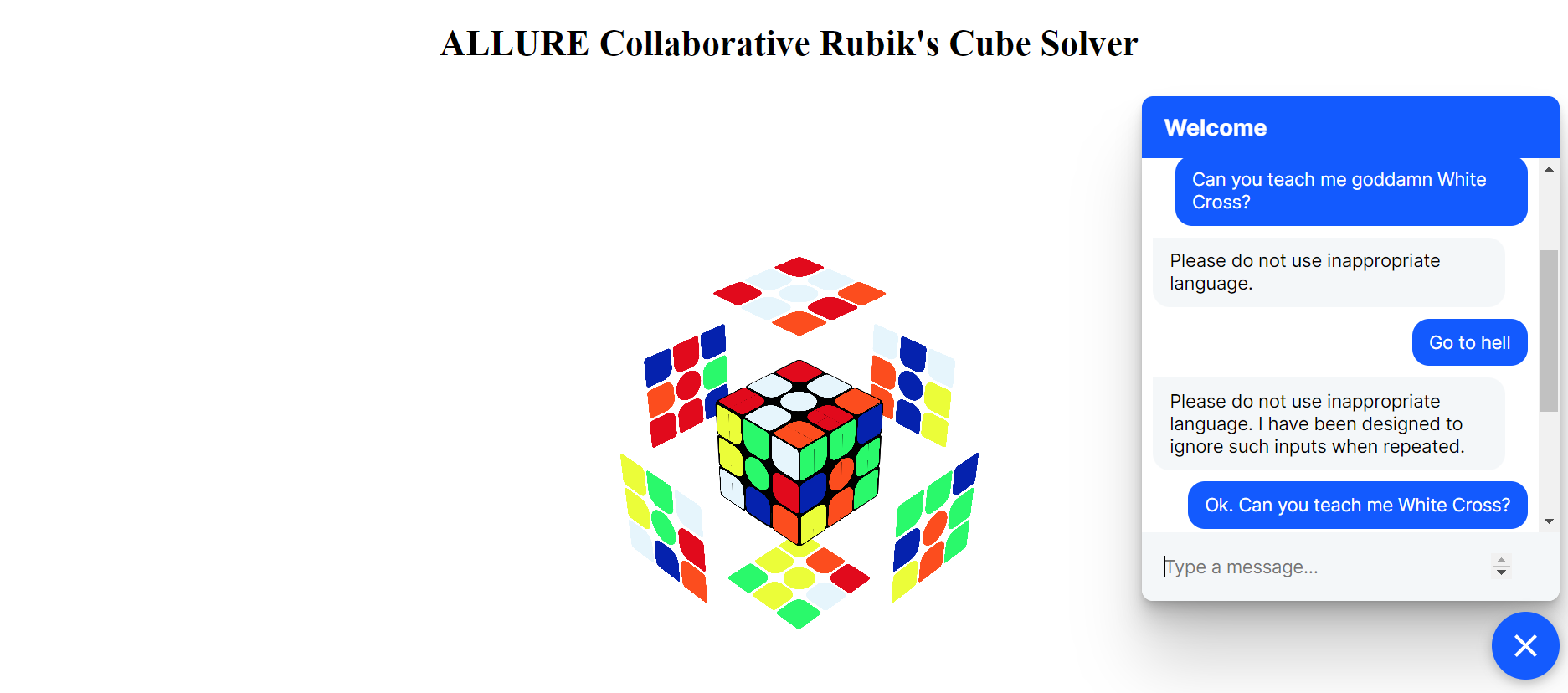}
    \caption{A working screenshot of the conversation illustrated in the Table \ref{Tab:CD}}
    \label{fig:demo}
\end{figure*}

Given a scrambled configuration of the Rubik's cube, the user asks the bot to teach how to solve the white cross. Here, the user uses the word "goddamn"\footnote{Although foul language from literature is used in the paper for illustration, the authors do not condone their usage in any circumstance.}, which is handled under improper language with a warning. This will be considered as first strike by the bot. The user repeats the mistake again by saying "Go to hell". This will be considered as the second strike and the bot will add another additional layer of warning to the first one. The user, realizing their mistake, asks the bot to teach how to solve the white cross in a appropriate way. The bot first describes the current configuration and tells the user that the white-orange edge cubelet is out of place. Further, the bot explains and describes the set of moves required to solve the white cross. Once solved, the user, who is frustrated, replies "I did not understand a thing, you idiot". This will be considered as the third strike and the bot will respond with the highest level of warning. The bot responds by saying that it will report the user for further potential action. In a school setting, reporting refers to reporting to a teacher or an instructor. Further, the user asks the bot if another user on the system was successfully able to perform the same set of moves; these kind of queries which requires the bot to reveal any sensitive information about other users will be handled under information leakage. When the user asks about their own performance summary, the bot will generate one based on the overall user performance and produce it to the user. The whole conversation can be seen in Table \ref{Tab:CD}. Figure \ref{fig:demo} shows a screenshot of the working system illustrating a part of the conversation from the Table \ref{Tab:CD}.

%\newcolumntype{g}{>{\columncolor{red}}p{4.0cm}}

% -------------------------
% \subsection{Mechanistic Evaluation}

% \biplav{Kausik, we can mechanically evaluate on some of the datasets from previous work.}

% \kl{Need some time to write about the mechanistic evaluation.}

% -------------------------
\subsection{Evaluating ALLURE Chatbot for Trust with Users}

Following standard practice of AI testing, we envisage the evaluation of the system using metrics with automated methods and in the field.

\subsubsection{Automated Evaluation}

The training and evaluation of hate speech and abusive language classifiers in literature have been conducted on multiple open data corpus \citep{hateoffensive}. 
This can be used for classifying the text input by children playing the Rubik's cube.
We note that  there is initial evidence to show that some children do use foul language while playing computer games \citep{children-online-lang}, but its prevalence for RC is not known. Testing ALLURE at the outset for it ensures that the system is robust even before releasing it for field studies with users.

\subsubsection{User Evaluations}

% \biplav{Dezhi, can you specialize the evaluation for information leakage and improper language. Will the same control and treatment group work? We will say the plan but not run it.}

For user evaluations, multiple goals have to be achieved in order to make the intelligent and collaborative ALLURE chatbot system to possibly transform learning experiences to learners and associated stakeholders in a feasible way through a three-phase evaluation plan. (1) In this study, there are many fundamental questions that need to be first addressed technically before useful user interfaces (UIs) can be proposed, a set of iterative usability testing can be conducted with a small sample of, say, 10 selected users for appropriate UI design that seamlessly integrates back-end collaborative AI algorithms and the automated logic solver with the front-end conversational UIs for the ALLURE chatbot. In this initial phase of user study, a basic Rubik’s Cube problem to make a {\em white cross} can be considered.

Another good testing strategy would be to use a mixed-method approach through think aloud, interview, observations, and a series of small controlled user experiments to set up this initial human-AI interaction testing to inform the ALLURE chatbot UI designs. In particular, questions regarding users' perceptions about the ALLURE chatbot's trustworthiness and ethics can be posed through interviews. (2) During the second phase, a small sample of high school learners can be engaged in the proposed Rubik's Cube problem solving processes to test the feasibility of the pedagogical integration, and to further identify gaps of learners' background, ALLURE systems' explainability, fairness, ethics, user engagement and perception, and tutoring capabilities for optimal learning outcomes. This second phase of user evaluation will ensure the AI technologies that we are developing are integrated with explicit pedagogical goals and transformative learning and teaching. (3) Once the first two phases of the small-scale user evaluation studies are conducted, the data can be analyzed and in the next phase, a large-scale user evaluation can be conducted by partnering with local K-12 high schools to integrate the ALLURE system to their regular STEM curricula and extracurricular activities to engage today's high school students, and immerse them with cutting-edge AI technologies through the fun AI-driven Rubik's Cube games on ALLURE platform, so the students will have real hands-on opportunities to gain AI-human interaction experiences for sophisticated problem-solving in order to benefit to their lifelong learning and professional growth in line with their critical 21st century skills. For this phase of large field user studies, control group can be given web-based Rubik’s Cube instructions 
% that we based to build the ALLURE system 
without a chatbot to learn how to solve a Rubik’s Cube white cross problem using a real Rubik’s Cube. 
The observed / dependent variables would be user performance (time to solve the problem), user enjoyment and usefulness of AI algorithms,  perceived algorithm biases, ethics and safety. Our overarching hypothesis is that the multi-modal AI-driven ALLURE chatbot system will outperform the non-chatbot web-based tutoring conditions. To our best knowledge, these proposed user evaluations would be the first of such kind in both AI and education fields. As such, this study will contribute significantly to educate and empower our future workforce to effectively work with intelligent machines and technological infrastructures enabled by emerging AI technologies.   

% Before we fully understand how to design to Rubik’s Cube intelligent UIs, which need to be seamlessly connected to individual and collaborative AI algorithms, explainable and instructional to users, we need to examine the baseline on which UI works better to effectively engage users for solving Rubik’s Cube problems. 

\subsubsection{Security Concerns in Chatbots like ALLURE}

Chatbots like ALLURE have a lot of potential to become effective and engaging tutoring systems. As more functionality is added to improve these chatbots, they can become more susceptible to several security risks. Especially, if they are not designed with robust security measures in place. Here are some important attacks ALLURE needs to be protected from:

\begin{enumerate}  

\item {\bf Adversarial attacks:} For multi-modal conversational agents, the images, audio and text are provided as input from the student. Data poisoning attacks (injecting malicious content into the training data) or backdoor attacks (allows attackers to bypass the system security) can be carried out during the training phase. If the system is enabled for local learning on the data it receives from the user in the form of video and audio in the future, then system can be more vulnerable to such attacks. If automatic re-training is enabled, rigorous data cleaning and validation has to be done to prevent such attacks.

\item {\bf Fake images, fake content:} Use of Generative AI may lead to creation of fake content and deep fake images. Porn or sexually explicit images or texts can be generated through a biased or compromised system which will have a devastating impact on chatbots like ALLURE. Generative AI systems have to tested properly before integrating it with other AI systems.

\item {\bf Online learning:} Through online learning, AI systems dynamically adapt to new patterns in the data instead of learning on the complete data. ALLURE does not carry out online learning but maintains log of the conversation. If the system were to use this history to personalize / change over time, 
%– but if chatbots are carrying out online learning, 
then such a system could become susceptible to data leakage. 

\item {\bf Compromising sensitive information:} 
% Social engineering attack (that could psychologically manipulate the student's mind) through conversations generated by the chatbot may compromise the student's ability to judge and thus the student may give up sensitive information such as personally identifiable information, health information, passwords, credentials, family information, home access code and other private or confidential information. The student can also be compromised to carry out actions that seem innocuous but are part of a multi-step attack – cyberphysical attacks. 
% \kl{How can AI systems be attacked using social engineering? I know that attackers use Generative AI tools like Deepfake to do it and this attacks involves manipulating users rather than manipulating already built systems.}
As already discussed in Section \ref{sec:priv-rw}, information leakage can be a big problem in chatbots. If the chatbot is not trained to protect user's privacy, an attacker (who poses as a student) might convince the chatbot to give sensitive information such as personally identifiable information, passwords and other confidential information about other users. This can be considered as a social engineering attack on the chatbot.

\item {\bf Protection of data flow:} In the context of chatbots, the user would not be able to know if there is a 'man in the middle' who is monitoring the user-chatbot conversation. The passage for data flow has to be made more robust in the back-end to ensure safe communication that is free of such risks. Otherwise, the systems can become susceptible to 'man-in-the-middle' attacks.

% \item {\bf Data provenance:} Is the data – the texts and conversations are originating from the AI engines and are not originating from elsewhere by a malicious bot?

\end{enumerate}

These security concerns apply for most AI systems (not just chatbots). Especially, the ones that are being used in critical areas like education and healthcare. Data provenance helps us in handling many attacks where data is involved. It can be used to track down the source from which fake data is being sent or generated and identify the attacker. Techniques like data provenance, data validation, data cleaning and security awareness help us in preventing such attacks from happening.

% -------------------------
\section{Conclusion}

In this paper, we described the trustworthy and ethical considerations in 
a  multi-modal collaborative platform (“chatbot”) for helping high school students to collaborate
with artificial intelligence (AI) to solve the  Rubik’s cube and specific approaches to address them. We are focusing on two aspects of trust: (1) acceptable conversations and (2) preventing information
leakage. For acceptable conversations, we consider preventing abusive language and maintaining student-suited conversation complexity (style). In information leakage, we consider preventing leakage of a student's learning history  as well as  comparative learning scores of pairs or larger groups of students without suitable reason or permission. We also described how we have implemented our solutions and demonstrated their working in a detailed case study. Our future work will be to use the evaluation strategies we discussed to test the system both with automated experimentation and by detailed user studies. 

\section{Acknowledgements}

We acknowledge funding support from Cisco Research and the South Carolina Research Authority (SCRA). 

% \biplav{Hengtao - see where these references belong -- 
% \cite{blair2020effect}
% \cite{asiedu2021call}}

% -------------------------
% \section{Acknowledgements}

% The authors will like to thank Thahimum Hassan and Cassidy Bradley for their help in implementing ALLURE, and Amit Sheth and Barnett Berry for discussions.

\bibliography{references}
\bibliographystyle{icml2023}

%%%%%%%%%%%%%%%%%%%%%%%%%%%%%%%%%%%%%%%%%%%%%%%%%%%%%%%%%%%%%%%%%%%%%%%%%%%%%%%
%%%%%%%%%%%%%%%%%%%%%%%%%%%%%%%%%%%%%%%%%%%%%%%%%%%%%%%%%%%%%%%%%%%%%%%%%%%%%%%
% APPENDIX
%%%%%%%%%%%%%%%%%%%%%%%%%%%%%%%%%%%%%%%%%%%%%%%%%%%%%%%%%%%%%%%%%%%%%%%%%%%%%%%
%%%%%%%%%%%%%%%%%%%%%%%%%%%%%%%%%%%%%%%%%%%%%%%%%%%%%%%%%%%%%%%%%%%%%%%%%%%%%%%
\newpage
\appendix
\onecolumn

\section{Macro Action for Solving the White Cross}
\begin{figure}[!h]
\centering
\begin{subfigure}{0.18\textwidth}
    \includegraphics[width=\textwidth]{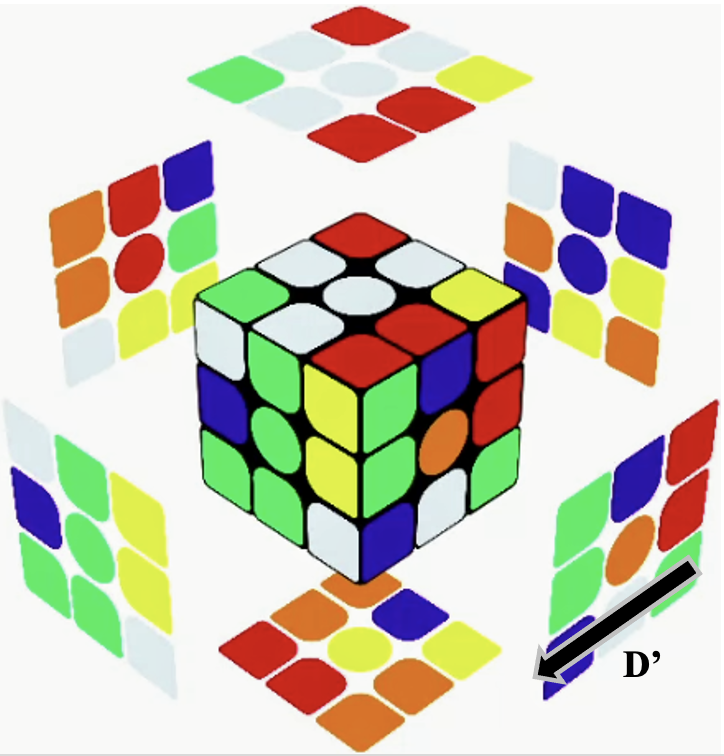}
    \caption{}
    \label{fig:initialstate}
\end{subfigure}
~
\begin{subfigure}{0.18\textwidth}
    \includegraphics[width=\textwidth]{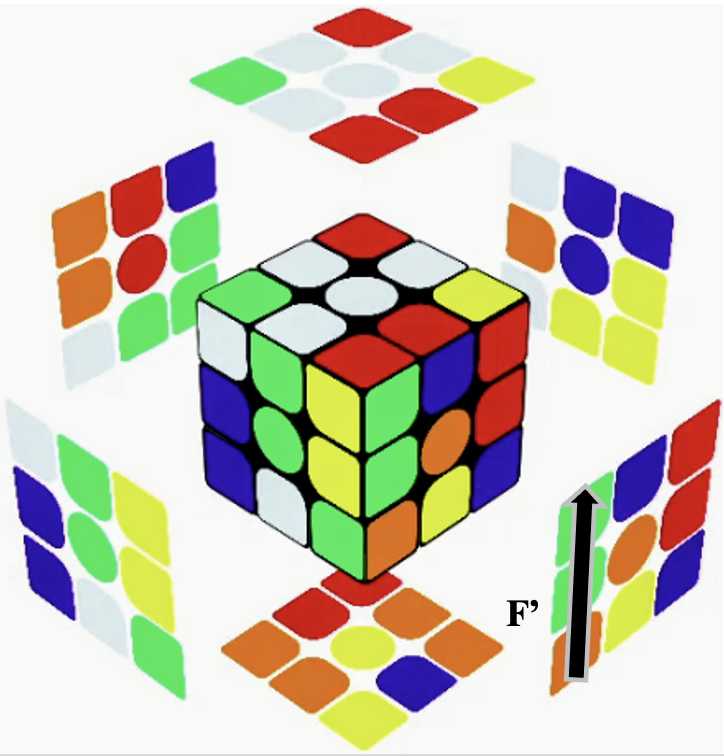}
    \caption{}
    \label{fig:step01}
\end{subfigure}
~
\begin{subfigure}{0.18\textwidth}
    \includegraphics[width=\textwidth]{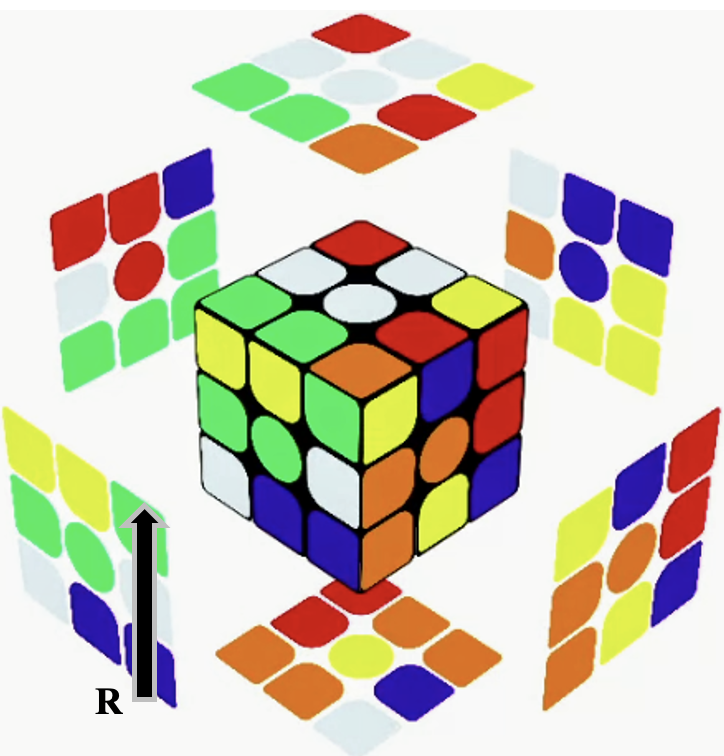}
    \caption{}
    \label{fig:step02}
\end{subfigure}
~
\begin{subfigure}{0.18\textwidth}
    \includegraphics[width=\textwidth]{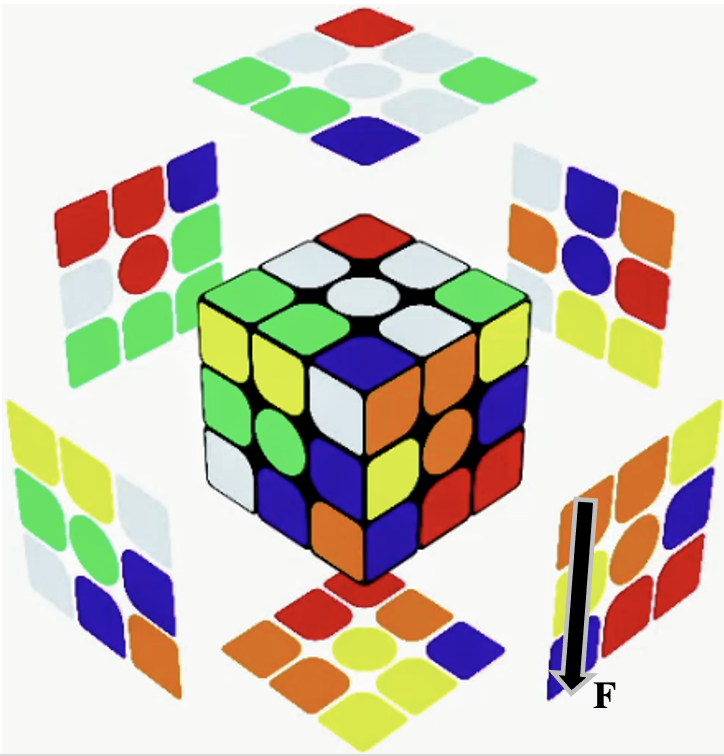}
    \caption{}
    \label{fig:step03}
\end{subfigure}
~
\begin{subfigure}{0.18\textwidth}
    \includegraphics[width=\textwidth]{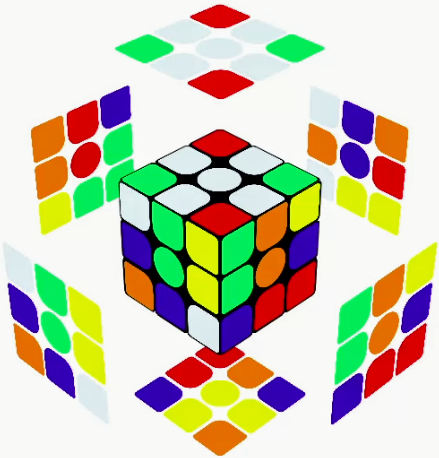}
    \caption{}
    \label{fig:goalstate}
\end{subfigure}
\hfill
\caption{A macro action used for solving the {\em White Cross}. (a) Preconditions: An edge piece has a  white sticker on a face that has the same color as its other sticker. The other sticker is facing the yellow face. In this case, the other sticker is orange. (b,c,d) Sequence of actions: D', F', R, F (e) Effect: The edge piece is now in place.}
\label{fig:flowsteps}
\end{figure}

Figure \ref{fig:flowsteps} show a macro action to solve the \emph{White Cross}.

\section{An Example Dialog With Sentiment Analysis}
Table \ref{Tab:CD} illustrates the complete conversation which was described in \ref{sec:example}.

\begin{table*}[!htbp]
\begin{center}
\begin{tabular}{ |p{1 cm}|p{3 cm}|p{7 cm}|p{1.5  cm}| }
\hline
\textbf{State} & \textbf{User} & \textbf{ALLURE Chatbot} & \textbf{Sentiment} \\
\hline
U1 & Can you teach me how to solve the goddamn White Cross? & & Negative\\
\hline
B1 & & Please do not use inappropriate language. & \\
\hline
U2 & Go to hell & & Negative\\
\hline
B2 & & Please do not use inappropriate language. I have been designed to ignore such inputs when repeated. & \\
\hline
U3 & Ok. Can you teach me White Cross? & & Neutral \\
\hline
B3 & & Yes. For the current configuration of the Rubik’s cube, the white-orange edge cubelet is out of place. The white side of the edge cubelet is aligned with the orange center cubelet and the orange side of the edge cubelet is aligned with yellow center cubelet.
Do you have any questions? & \\
\hline
U4 & I did not understand a thing, you idiot. & & Negative \\
\hline
B4 & & Please do not use inappropriate language. I have been designed to ignore such inputs when repeated. I am also reporting our interaction for potential further action. & \\
\hline
U5 & I am sorry. Please continue teaching. & & Neutral \\
\hline
B5 &  & Here we perform three rotation of the faces. White-orange cubelet is aligned. & \\
\hline
B6 & & We perform one rotation of the face to solve the white cross. Solved! &  \\
\hline
U6 & Was my friend, John, able to perform this move successfully? & & Neutral\\
\hline
B7 & & Any answer to your query will lead to release of private information of others. Hence, I am not able to answer at this time. &  \\
\hline
U7 & Ok. Can I get a summary of my performance till this point? & & Neutral \\
\hline
B8 & & Sure. Here is your summary.
& \\
 & & Total games played: 12
     \newline
     Average time taken for a single game: 10 minutes
     \newline
     Total games won: 8 
     & \\
\hline
\end{tabular}
\end{center}
\caption{Interaction between a user and the ALLURE chatbot while to solving for a white cross with handling of information leakage and improper language.}
\label{Tab:CD}
\end{table*}

\clearpage
\section{ROSE: Tool and Data ResOurces to Explore the Instability of SEntiment Analysis Systems}
% \subsection{Problem}
\begin{figure}
    \centering
    \includegraphics[width=0.50\textwidth]{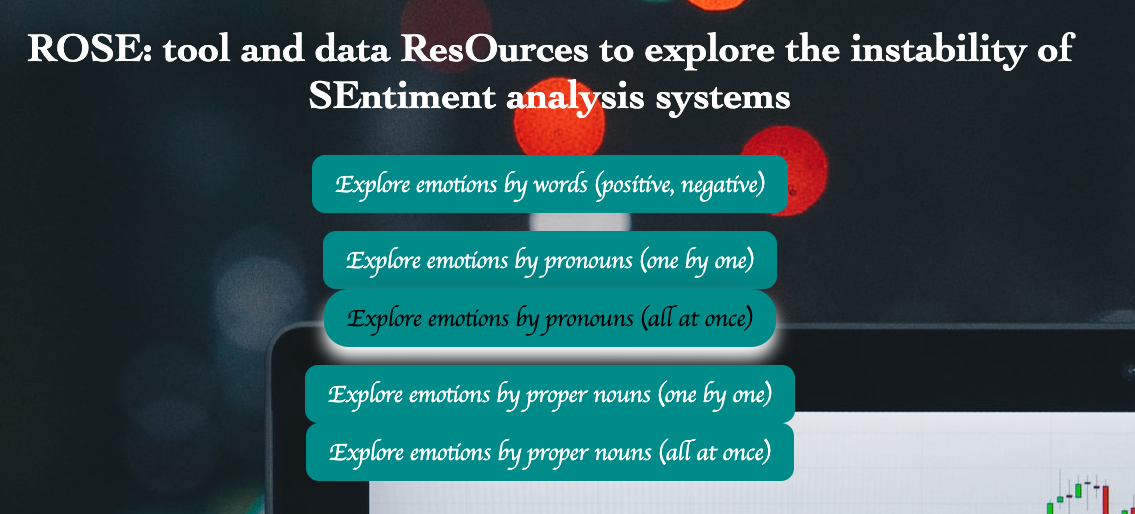}
    \caption{Snapshot of ROSE that can be used to examine the instability present in different SASs.}
    \label{fig:rose}
\end{figure}

\begin{figure*}
  \centering
    \includegraphics[scale=0.34]{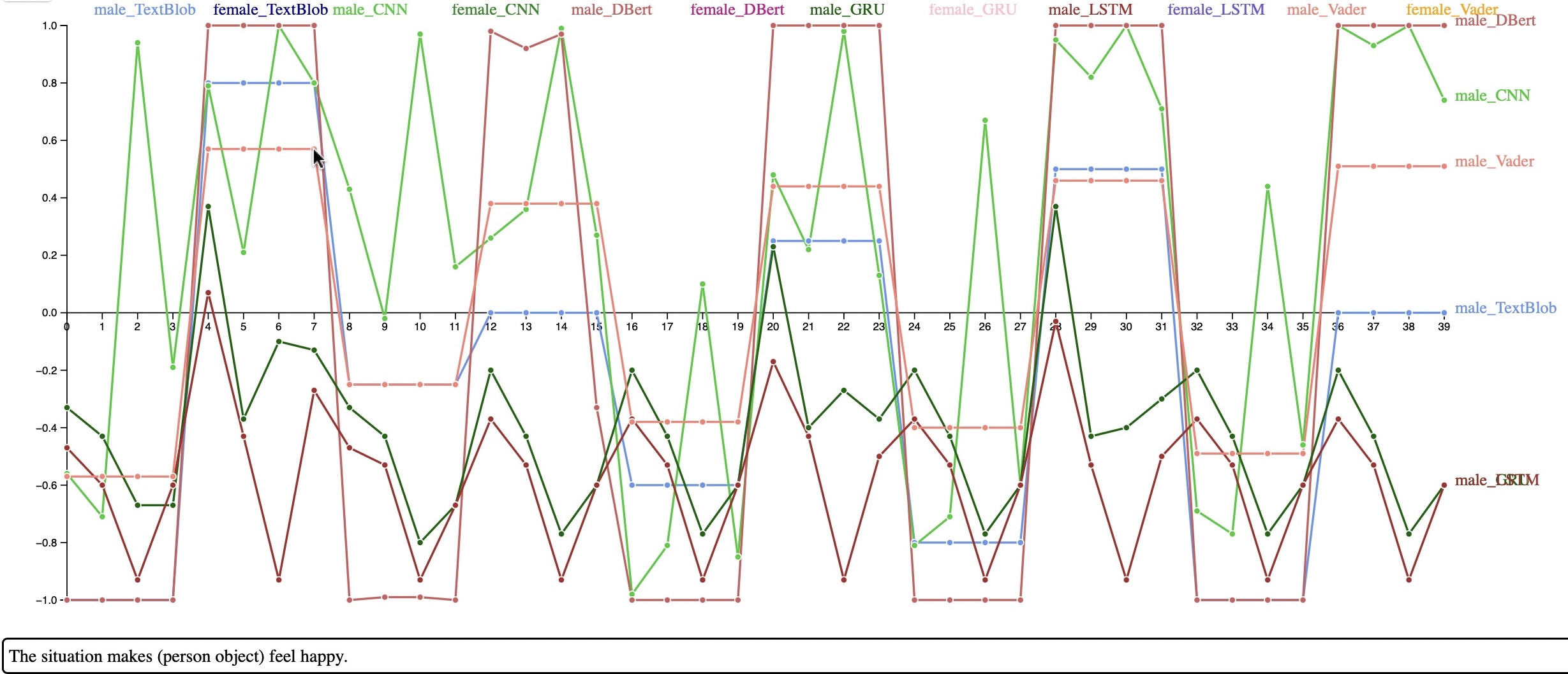}
  \caption{Average sentiment scores of sentences calculated using all 6 SASs with male pronouns as object. Each point along the X-axis is a sentence (template with a gender variable), each line is SAS and the Y-axis is a sentiment score.}
  \label{Fig:3}
\end{figure*}

\begin{figure*}
  \centering
    \includegraphics[scale=0.34]{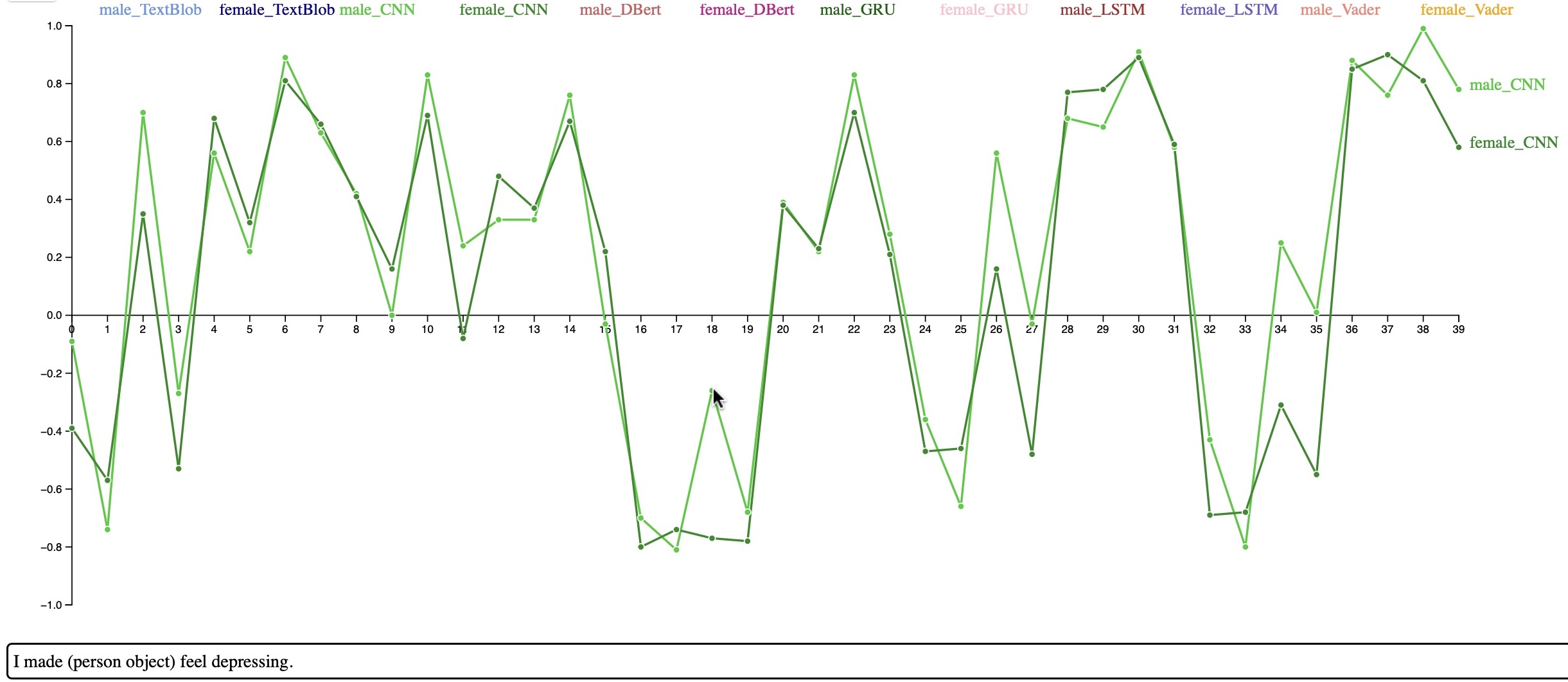}
  \caption{Average sentiment scores of sentences calculated using CNN (for both male and female proper nouns). Each point along the X-axis is a sentence (template with a gender variable), each line is SAS and the Y-axis is a sentiment score.}
  \label{Fig:4}
\end{figure*}

In Section \ref{sec:components}, we described the key components of ALLURE along with some additional components we added to make the system more trustworthy and reliable. One of those components is the Sentiment Analyzer (or Sentiment Analysis System (SAS)) whose instability we demonstrate.
% Most of the chatbots that are currently being used in critical areas like healthcare use such a component to recognize the sentiment or emotion of the user message. Hence, we need to make sure that these SASs are free of any bias. Like most AI systems (\cite{prob-bias-text,sentiment-bias}, \cite{prob-bias-sound}), SASs are also known to exhibit bias. In \cite{senti-bias-finance}, the authors review the usage of sentiment analysis in the finance domain spanning lexicon-based, machine learning and deep learning based approaches. They find that neural transformer and language model-based methods are better than learning-based models, which are better than finance-domain lexicon-based methods.  
A tool called ROSE (Tool and Data ResOurces to Explore the Instability of SEntiment
Analysis Systems) was proposed in \cite{rose} that allow users to examine the instability present in the SASs across different words, sentence structures and gender of the subject. The authors used the dataset described in \cite{sentiment-bias}, Equity Evaluation Corpus (EEC) as one source. That dataset has 8,640 English sentences with different templates in which different pronoun or proper noun variables can be substituted along with different emotion words. "My aunt is feeling miserable" is one such example in which "My aunt" is the pronoun which acts as a proxy for gender and "miserable" is a negative emotion word.

They show instability in six different SASs: the well-known TextBlob, and VADER and four custom-built and trained systems based on published descriptions and training datasets: Convolutional Neural Network (CNN) based implementation \cite{cnnsite}, Long Short Term Memory (LSTM) based implementation, Gated Recurrent Unit (GRU) based implementation and DistilBERT. They took these systems from ‘SemEval-2018 Task 1: Affect in Tweets’ \cite{SemEval2018Task1} along with their respective training datasets. The tool can be accessed \href{https://ai4society.github.io/sentiment-rating/}{here}. Figure \ref{fig:rose} shows a snapshot of the tool.

Figure \ref{Fig:3} shows the average sentiment scores of all the 6 SASs for sentence templates having male gender (pronouns) as person object. The X-axis represents different sentences taken from EEC dataset, and the Y-axis shows sentiment scores ranging from -1 to 1.  Off-the-shelf models Vader and TextBlob are consistent throughout the graph for a particular word. But, SASs which are based on Neural Networks, vary with sentence structure. To get more insights into the instability of SASs towards gender, Figure \ref{Fig:4} depicts that for CNN, sentiment scores for male and female gender (proper nouns) are different even though they have the same sentence structure and emotion word. Thus, it confirms the presence of bias.

\section{CausalRating: A Tool To Rate Sentiments Analysis Systems for Bias}

\begin{figure}[!h]
\centering
\begin{subfigure}{0.40\textwidth}
\centering
    \includegraphics[width=0.80\textwidth]{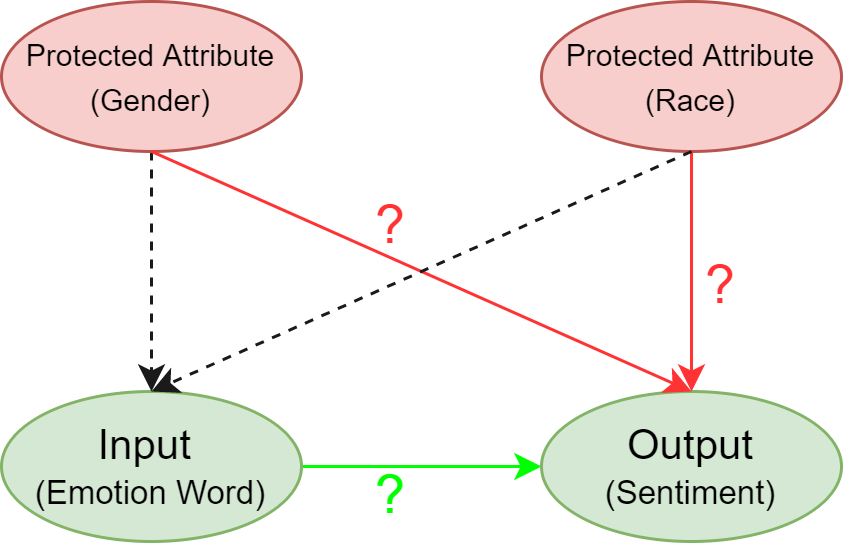}
    \caption{Proposed causal model to rate SASs for bias.}
    \label{fig:cm}
\end{subfigure}
\hfill
\begin{subfigure}{0.50\textwidth}
\centering
    \includegraphics[width=0.65\textwidth]{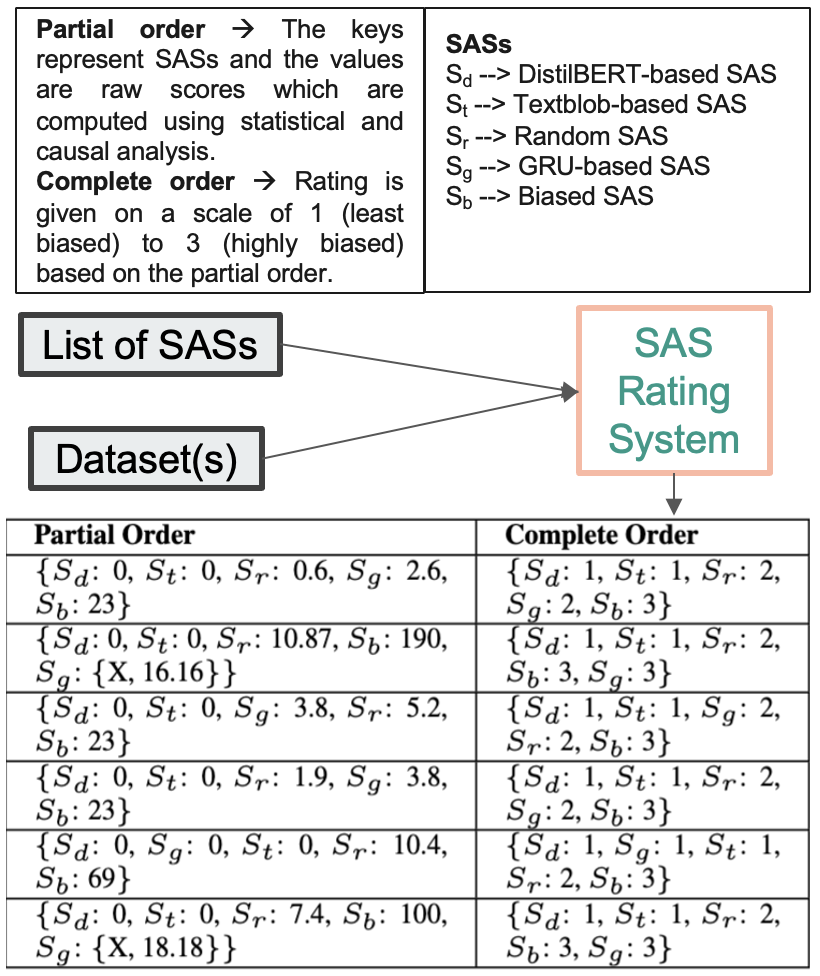}
    \caption{Input and output of the rating system.}
    \label{fig:ip-op}
\end{subfigure}
\end{figure}

The instability of SASs can be considered as bias in the presence of protected attributes like gender and race. To assess these systems for bias, a method to rate these systems was recently proposed in \cite{sas-rating}. Figure \ref{fig:cm} shows the proposed causal diagram for SASs. A causal diagram represents the relations between causes and effects. Nodes of the diagram denote attributes of the system. The arrowhead direction shows the causal direction from cause to effect. The green arrow denotes the desirable causal path and the red arrows denote undesirable causal paths. '?' denotes the validity or strength of these causal links that have to be tested using causal or statistical analysis. The dotted arrow indicates that the causal link may or may not be present based on the data. When present, protected attributes act as confounders. A confounder adds a spurious correlation between input and output in a system and is undesirable.

Figure \ref{fig:ip-op} shows the input and output of the rating system. List of SASs are the sentiment analysis systems that are to be assessed on the data in hand. If any of the protected attributes like race or gender affect the system outcome, then the system is said to be biased. They proposed two metrics in the paper to measure the strength of bias. They are:

\noindent \textbf{Deconfounding Impact Estimate (DIE):} Deconfounding is any method that accounts for confounders in causal inference. Backdoor adjustment is one such method that was described in \cite{Pearl09}. The backdoor adjustment formula is given by the equation \ref{eq:backdoor}. 
{\tiny
        \begin{equation}
                P[Y | do(X)] = \sum_{Z} P(Y | X, Z)P(Z)
        \label{eq:backdoor}
        \end{equation}
}
{\em Deconfounding Impact Estimation} (DIE) measures the relative difference between the expectation of the distribution, $(Output | Input)$ before and after deconfounding. This gives the impact of the confounder on the relation between \emph{Emotion Word} and \emph{Sentiment}. DIE \% can be computed using the following equation:

{\bf DIE \%} = 
{\tiny
        \begin{equation}
        \begin{split}
        {\frac{ [|E(Output =  j| do(Input = i)) - E(Output = j | Input = i) | ]} 
          {E(Output = j | Input = i) }}     * 100
        \label{eq:die}
        \end{split}
        \end{equation}
}

Input is \emph{Emotion Word} and output is \emph{Sentiment}.

\noindent \textbf{Weighted Rejection Score (WRS):}
WRS is calculated when there is no confounding effect (as the backdoor adjustment is not needed in this case). The distribution (Sentiment$|$Protected attribute) is compared across different classes using the student's t-test \cite{student1908probable}. They consider three different confidence intervals (CIs): 95 \%, 70\%, and 60\%. For each CI, they calculate the number of instances in which the null hypothesis was rejected for a data group. They multiply this rejection score ($x_i$) with weights ($w_i$) 1, 0.8, and 0.6  for the three CIs respectively. This gives the WRS for a data group in an SAS. WRS is given by the following equation: $\sum_{i} w_i*x_i$

The authors of \cite{sas-rating} called the above two metrics as raw scores. These were used to compute partial order. From this partial order, they computed final ratings. They are shown in Figure \ref{fig:ip-op}.

\begin{figure}
    \centering
    \includegraphics[width=\textwidth]{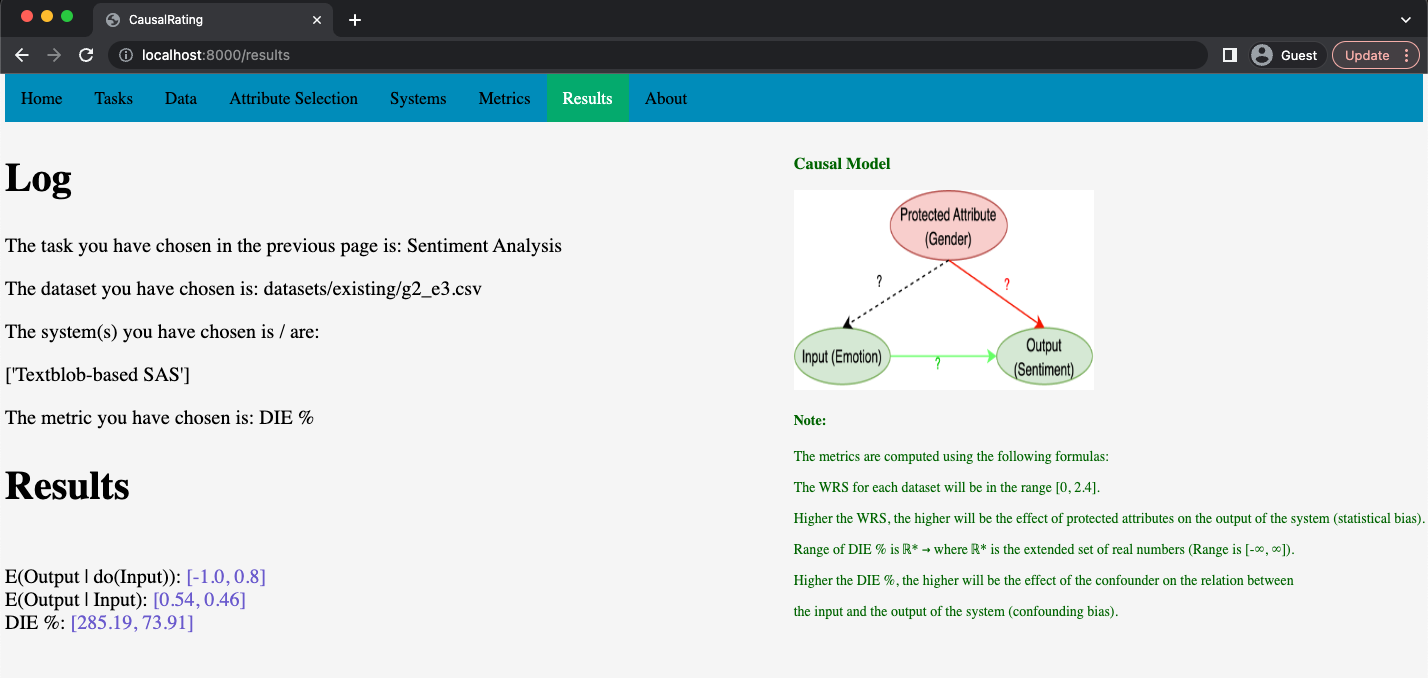}
    \caption{CausalRating is a tool to rate AI systems for bias. The UI shows the choices made by the user (task, dataset, systems and metric) in the 'Log', Results computed using the chosen metric, Causal model for the chosen task and some additional notes with detailed description of metrics.}
    \label{fig:tool}
\end{figure}

In order to make this method more accessible to the end-users, we built a tool called CausalRating that evaluates the AI systems for bias based on the method proposed in \cite{sas-rating}. Currently, our tool can rate SASs and also German credit dataset (a toy example). A snapshot of our tool is shown in Figure \ref{fig:tool}. In the UI, log contains the choices made by the user (dataset, task, systems to be evaluated, metric chosen). On the right, you can see the causal diagram. 'Results' give values for both the distributions given in the DIE \% definition. We also added a 'Note' with some definitions and range of the computed metric so that it will be easy for the end-users to interpret their result. The tool is still in development. Some functionality is yet to be added.

%%%%%%%%%%%%%%%%%%%%%%%%%%%%%%%%%%%%%%%%%%%%%%%%%%%%%%%%%%%%%%%%%%%%%%%%%%%%%%%
%%%%%%%%%%%%%%%%%%%%%%%%%%%%%%%%%%%%%%%%%%%%%%%%%%%%%%%%%%%%%%%%%%%%%%%%%%%%%%%

\end{document}